\documentclass[12pt]{article}

\usepackage{amsmath}
\usepackage{amssymb}

\newcommand{\tmop}[1]{\operatorname{#1}}
\newcommand{\beq
}{\begin{equation}}
\newcommand{
\eeq}{\end{equation}}
\newcommand {\thetab}{\bar{\theta}}
\newcommand {\ep}{\epsilon}
\newcommand{\zb}{\bar{z}}
\begin{document}
\begin{titlepage}
\begin{flushright}
YITP-SB-04-55  \\
{\tt hep-th/0410081}\\
\end{flushright}
\vskip 22mm
\begin{center}
{\huge {\bf On Calabi-Yau supermanifolds II}}
\vskip 10mm
{\bf Martin Ro\v{c}ek}\\
{\em C.N. Yang Institute for Theoretical Physics}\\
{\em SUNY, Stony Brook, NY 11794-3840, USA}\\ 
{\tt rocek@insti.physics.sunysb.edu}
\vskip 6mm
{\bf Neal Wadhwa}\\
{\em Ward Melville High School}
\end{center}
\vskip .2in

\begin{center} {\bf ABSTRACT } \end{center}
\begin{quotation}\noindent
We study when Calabi-Yau supermanifolds $\mathbb{M}^{1|2}$ with one complex
bosonic coordinate and two complex fermionic coordinates are super Ricci-flat, 
and find that if the bosonic manifold is compact, it must have constant scalar curvature.
\end{quotation}
\vfill
%%%%%%%%%%%%%%%%%%
\end{titlepage}
\newpage

In \cite{rw}, we found that super Ricci-flat K\"ahler manifolds with one fermionic 
dimension and an arbitrary number of bosonic dimensions exist above a bosonic 
manifold with a vanishing Ricci scalar. This paper explores super Calabi-Yau 
manifolds with one bosonic dimension and two fermionic dimensions. We find that
the condition that the supermetric is super Ricci-flat implies
several interesting constraints that are familiar from other 
contexts, including the field equation of the WZW-model on $AdS_3$. Locally, 
these constraints imply that the super K\"ahler potential has the form
\beq
\label{result}
K(z,\zb,\theta,\thetab)=K_0(z,\zb)+\sqrt{K_0(z,\zb),_{z\zb}}\theta^i\thetab^i
+\frac14\!\left(\ln\!\left[K_0(z,\zb),_{z\zb}\right]\right)\!,_{z\zb}(\theta^i\thetab^i)^2~,
\eeq
where $K_0(z,\zb)$ is the K\"ahler potential of the bosonic manifold. We find the 
further constraint that the scalar curvature of the bosonic manifold is harmonic; 
on a complete compact space, this implies that the scalar curvature is constant.

Consider the super K\"ahler potential $K$ of the manifold $\mathbb{M}^{1|2}$ 
with 1 bosonic coordinate and 2 fermionic coordinates: 
\beq
 K=f_0+if_1\theta^2+i\bar{f}_1\thetab^2+f_{i\bar{j}}\theta^i\thetab^j+f_2\theta^2\thetab^2
\eeq
We use the notation $\theta^2=\frac12 \ep_{ij}\theta^i\theta^j$, where $\ep_{ij}=-\ep_{ji}$.  
Since $\overline{\theta^i\theta^j}=\thetab^j\thetab^i$, the factor of $i$ is needed to 
make $K$ real.  The supermetric of this manifold is 

\beq
\left(\begin{array}{cc}
	\left.\begin{array}{cc}f_{0,z\zb}+if_{1,z\zb}\theta^2+i\bar{f}_{1,z\zb}\thetab^2\\  
	+f_{i\bar{j},z\zb}\theta^i\thetab^j+f_{2,z\zb}\theta^2\thetab^2\end{array}\right. &
	i\bar{f}_{1,z}\ep_{ij}\thetab^i+f_{i\bar{j},z}\theta^i\frac{1}{2}+f_{2,z}\theta^2\ep_{ij}\thetab^i\\ & \\
	if_{1,\zb}\ep_{ij}\theta^j+f_{i\bar{j},\zb}\thetab^j+\frac{1}{2}f_{2,\zb}\ep_{ij}\theta^j\thetab^2 &
	f_{i\bar{j}}+f_{2}\ep_{il}\theta^l\ep_{kj}\thetab^k\end{array}\right)~
\eeq
	
The superdeterminant can be set to 1 by a holomorphic coordinate transformation as 
described in \cite{rw}.  Looking at only the bosonic part of the superdeterminant yields  
\beq
\label{pureb}
 f_{0,z\zb}=\det(f_{i\bar{j}})
\eeq
Equating the coefficients of the purely holomorphic 
and purely anti-holomorphic fermions gives the equation 
\beq
 \frac{f_{1,z\zb}}{f_{1,\zb}}=\frac{\det(f_{i\bar{j}})_{\zb}} {f_{0,z\zb}} 
\eeq 
as well as its conjugate. This is equivalent to $ \ln(f_{1,\zb})_z=\ln(\det(f_{i\bar{j}}))_{z}$ or 
\beq
 f_{1,\zb}=\lambda(\zb) \det(f_{i\bar{j}})
\eeq 
where $\lambda(\zb)$ is an arbitrary function of $\zb$. This implies that locally  $f_{1}$ 
can always be removed by a holomorphic coordinate transformation: If $\lambda=0$, 
$f_1=f_1(z)$ is holomorphic and contributes a holomorphic term to the super K\"ahler 
potential that does not change the supermetric. Otherwise, the coordinate transformation 
$\frac{1}{\lambda(\zb)}\frac{\partial}{\partial \zb}\rightarrow \frac{\partial}{\partial \zb'}$, 
results in the equation 
\beq
 f_{1,\zb}= \det(f_{i\bar{j}})= f_{0,z\zb}~,
\eeq 
where we have used (\ref{pureb}) in the last step.
Then the coordinate transformation, $z+i\theta^2\to z$ and 
$\zb+i\thetab^2 \to \zb$ eliminates $f_1$ (up to purely holomorphic terms), as can be verified by a 
Taylor expansion. This coordinate transformation is only valid locally, as $z$ 
and $\zb$ are not globally defined functions on all such manifolds. 
	
Thus the K\"ahler potential can be assumed to have the form: 
\beq
\label{simp}
 K=f_0+f_{i\bar{j}}\theta^i\thetab^j+f_2\theta^2\thetab^2 
\eeq 
and the simplified supermetric is 
\beq
\left(\begin{array}{cc}
	f_{0,z\zb}+ f_{i\bar{j},z\zb}\theta^i\thetab^j+f_{2,z\zb}\theta^2\thetab^2 &
	f_{i\bar{j},z}\theta^i+\frac12 f_{2,z}\theta^2\ep_{ij}\thetab^i\\ & \\
	f_{i\bar{j},\zb}\thetab^j+\frac{1}{2}f_{2,\zb}\ep_{ij}\theta^j\thetab^2 &
	f_{i\bar{j}}+f_{2}\ep_{il}\theta^l\ep_{kj}\thetab^k\end{array}\right)~.
\eeq
By looking at the coefficient of the $\theta\thetab$ terms, one can find that 
\beq
 \label{eq:mat}
 \frac{f_{b\bar{c}}f_{i\bar{a},z}f_{d\bar{j},\zb}\ep^{db}\ep^{ca}}
 {\tmop{det}(f_{i\bar{j}})}+f_{i\bar{j},z\zb}=-f_2f_{i\bar{j}} ~,
\eeq
where $\ep_{ab}\ep^{bc}=\delta_a{}^c$.
Multiplying by the inverse $f_{i\bar{j}}$ and contracting yields the equation 
\beq
 f_2=-\left(\ln\sqrt{\tmop{det}(f_{i\bar{j}})}\,\right)_{z\zb} ~.
\eeq
Substituting this into equation (\ref{eq:mat}) gives
\beq
\label{eq:ald}
(f_{i\bar{j},z} f^{\bar{j}k})_{\zb}
= \left(\tmop{ln}\sqrt{\tmop{det}(f_{i\bar{j}})}\,\right)_{z\zb}\delta_i{}^k~,
\eeq
where $f^{\bar i j}$ is the inverse of $f_{i\bar{j}}$.
Now let $M:=\frac{f_{i\bar{j}}}{\sqrt{\tmop{det}(f_{i\bar{j}})}}$; then 
$M$ is a hermitian matrix whose determinant is 1. It can be written as 
\beq
 M=\left(\begin{array}{cc} x+y &u+iv \\ u-iv & x-y\end{array}\right) ~.
\eeq 
The condition $\tmop{det}( M) =1$ implies $x^2-y^2-u^2-v^2=1$, which is a 
hyperboloid, and is well known in the physics literature as $AdS_3$ (see, 
{\it e.g.}, \cite{B}, where this space is used to study Black holes).

Equation (\ref{eq:ald}) implies 
\beq
 \label{eq:phy} 
 (M^{-1} M_{\zb})_z=0 \Leftrightarrow (M_z M^{-1})_{\zb}=0 ~.
\eeq 
This matrix differential equation is well studied and appears in many physical systems--
it is the classical equation of motion of the WZW-model on $AdS_3$ \cite{giveon}.
If we use the parameterization
\beq
M=\left(\begin{array}{cc} e^{-\phi}+
\gamma\bar{\gamma}e^\phi & \gamma e^\phi \\ \bar{\gamma}e^\phi & e^\phi\end{array}\right)~, 
\eeq 
then the four resulting equations give the functional gradient of equation (2.9) in \cite{giveon}. 

Equation (\ref{eq:phy}) implies that 
\beq
M=\mathcal{M}(z)Y\bar{\mathcal{M}}(\zb) ~,
\eeq
where $\mathcal{M}(z)$ is a holomorphic matrix and $\bar{\mathcal{M}}$ is its adjoint.
In this context, it is possible to go further and (locally) eliminate $M$ by applying 
the coordinate transformation $\theta \mathcal{M} Y^{\frac12}\rightarrow \theta$ and 
$Y^{\frac12}\bar{\mathcal{M}} \thetab \rightarrow \thetab$. This simplifies the super 
K\"ahler potential further to the form (\ref{result}).

The remaining terms in the superdeterminant, once one assumes 
$M_{i\bar j}=\delta_{i\bar j}$, imply
\beq
\label{eq:rres}R,_{z\zb}=0~, 
\eeq 
where $R$ is the Ricci scalar of the bosonic part. On a complete compact manifold, 
the only bosonic functions that obey (\ref{eq:rres}) are constant functions. This proves 
that the Ricci scalar is a constant on this type of manifold.

On a noncompact manifold, there are other solutions which may prove to be interesting; the
results of  \cite{dey} relate closely to equation (\ref{eq:rres}), and potentially could be useful
in pursuing this further.

An example of a space obeying all the conditions that we have found is $\mathbb{SP}^{^{1|2}}$. 
This space is compact and satisfies all of the constraints that we have found, guaranteeing that
its supermetric is super Ricci-flat. Its K\"ahler potential is 
\beq
\ln(1+z\zb)+\frac{\theta^1\thetab^1+\theta^2\thetab^2}{1+z\zb}
+\frac{\theta^1\theta^2\thetab^1\thetab^2}{(1+z\zb)^2} ~,
\eeq
The Ricci scalar of the bosonic part is 2 and $\ln(1+z\zb)_{,z\zb}=1/{(1+z\zb)^2}$. 
Another example is given by a Riemann surface $\Sigma$ with a constant negative Ricci scalar.

In this paper, many coordinate transformations are used that could be obstructed globally, {\it e.g.}, 
it may not be possible to remove the $\theta^2$ and $\thetab^2$ terms as well to transform $M$ to
the identity matrix; it would be interesting to see how such terms modify (\ref{eq:rres}) and if
other interesting solutions arise.

\noindent{{\bf Acknowledgement}:  We are happy to thank Leon Takhtajan for useful discussions 
and Rikard von Unge for comments on the manuscript. 
The work of MR was supported in part by NSF grant no.~PHY-0354776.}

\noindent{{\bf Note}: While writing up our results after completing our calculations, 
we became aware of \cite{z}, which has considerable overlap with our work. 
It studies the case with an arbitrary number of bosonic dimensions, but with 
a super K\"ahler potential assumed to have the simple form (\ref{simp}).
Because of the greater complexity of bosonic manifolds in higher dimensions, \cite{z} does not
find as complete results as those presented here.}

\end{document}